\documentclass{article}
\usepackage[utf8]{inputenc}

\usepackage{float}
\usepackage{graphicx}
\usepackage{subfig}
\usepackage{epstopdf}
\usepackage[format=plain,
            labelfont={bf},
            textfont=it]{caption}

\title{Re-annotation of cough events \\ in the AMI corpus}
\author{Paul Leamy, Ted Burke, Damon Berry, David Dorran \\ \\
\textit{Biomedical Research Group} \\
\textit{Technological University Dublin}}
\date{17-18 June 2019}

\begin{document}

\maketitle

\section{Introduction}
Cough sounds act as an important indicator of an individual's physical health \cite{el2011survey}, often used by medical professionals in diagnosing a patient's ailments \cite{schappert2006ambulatory}. 
In recent years progress has been made in the area of automatically detecting cough events and, in certain cases, automatically identifying the ailment associated with a particular cough sound.
Ethical and sensitivity issues associated with audio recordings of coughs make it more difficult for this data to be made publicly available. 
Without the public availability of a reliable database of cough sounds, developments in the area of audio event detection are likely to be hampered \cite{7472920}.
The purpose of this paper is to spread awareness of a database \cite{carletta2007unleashing} containing a large amount of naturally occurring cough sounds that can be used for the implementation, evaluation, and comparison of new machine learning algorithms that allow for audio event detection associated with cough sounds.
Using a purpose built GUI designed in MATLAB, a re-annotated version of the Augmented Multi-party Interaction (AMI) corpus' cough location annotations was produced with 1369 individual cough events. 
All cough annotations and the re-annotation tool are available for download and public use \cite{leamy_2019}.

\section{The AMI corpus}
The AMI corpus data is publicly available under the Creative Commons Attribution 4.0 license agreement, and contains audio recordings of indoor meetings taken in an office environment. 

\begin{figure}[h!]
    \centering
    \includegraphics[width=0.7\linewidth]{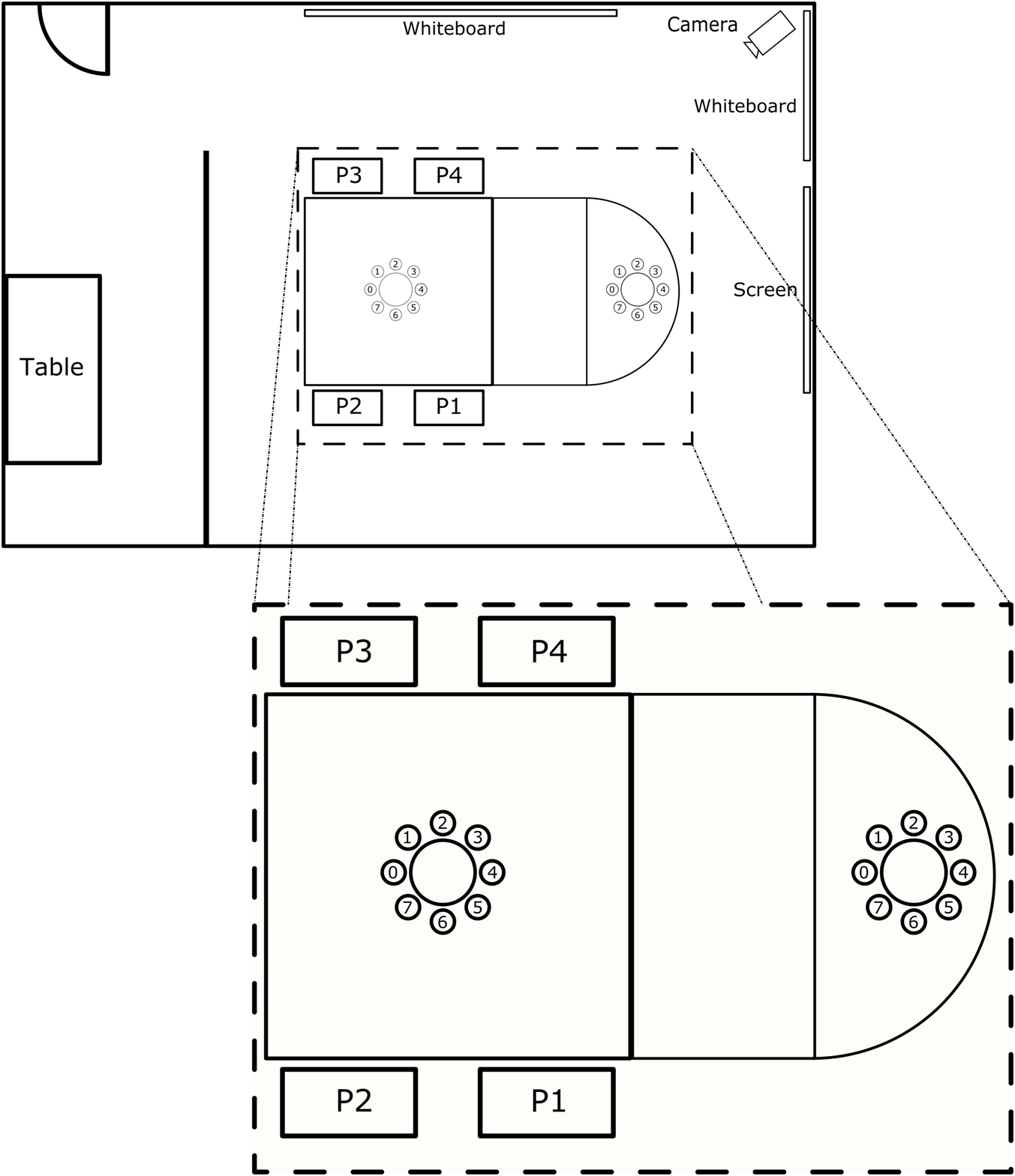}
    \caption{Recordings were captured from omni-directional microphone arrays (near and far mounted) and headset and lapel microphones - \textbf{Dashed insert:} Closeup of omni-directional microphone arrays.}
    \label{fig:RoomLayout}
\end{figure}

The audio recordings are accompanied by annotations of a large number of audio events throughout the entire corpus.
Audio recordings were captured from multiple microphone locations, with an example of one meeting room shown in a Figure \ref{fig:RoomLayout}.

\subsection{Re-annotation procedure}
\begin{figure}[h] 
    \centering
    \subfloat{\includegraphics[width=0.5\linewidth]{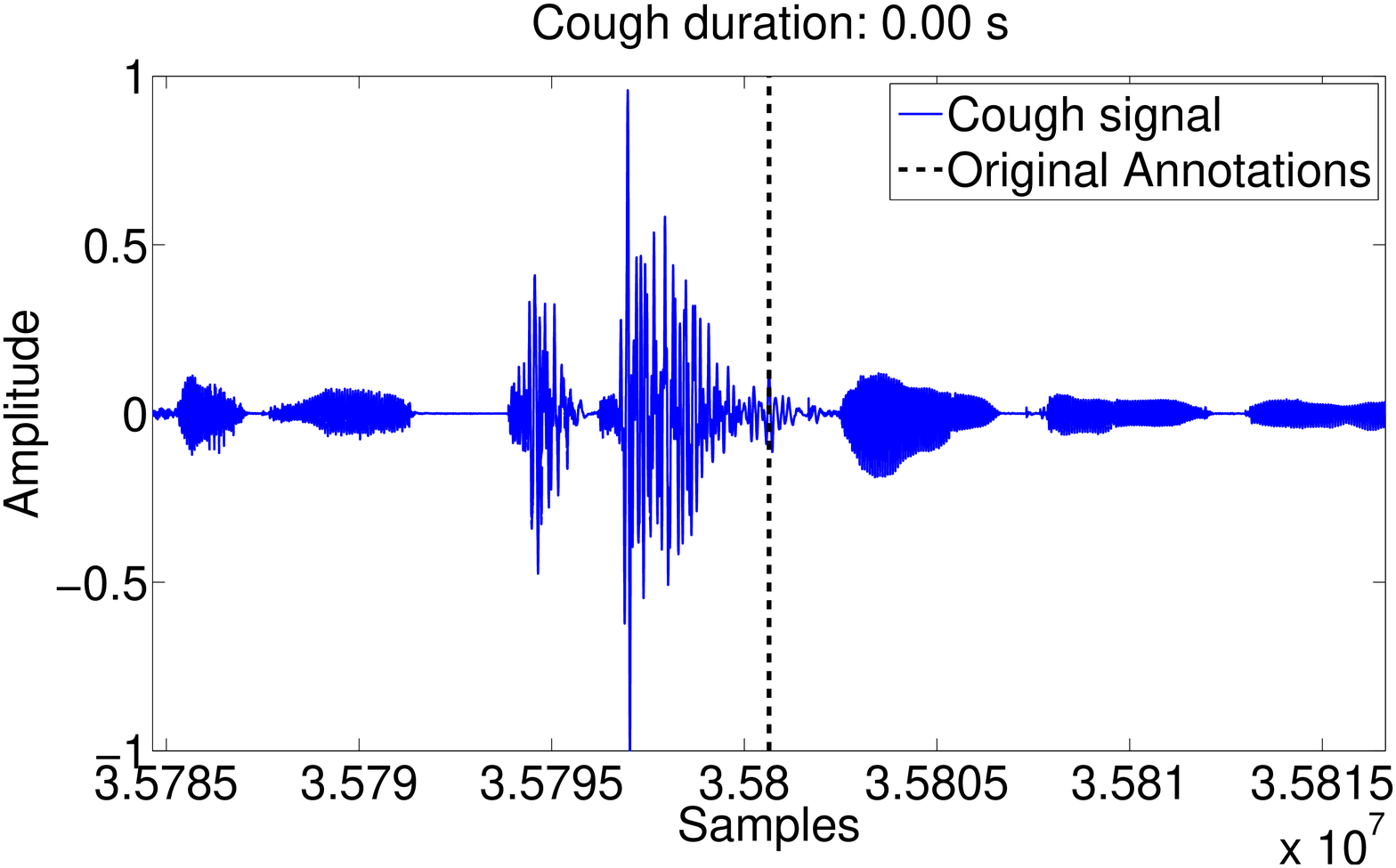}} 
    \subfloat{\includegraphics[width=0.5\linewidth]{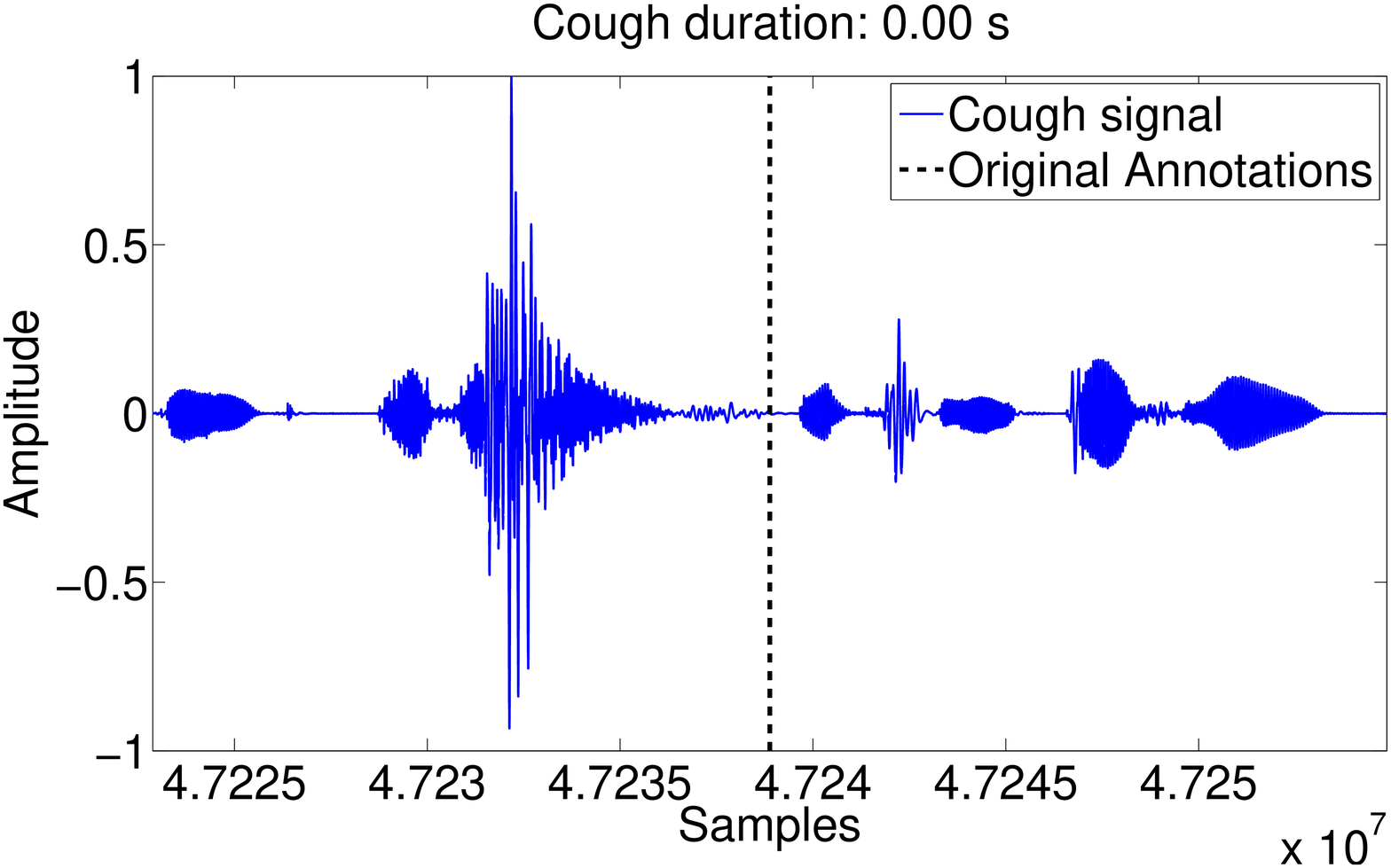}}\\
    \subfloat{\includegraphics[width=0.5\linewidth]{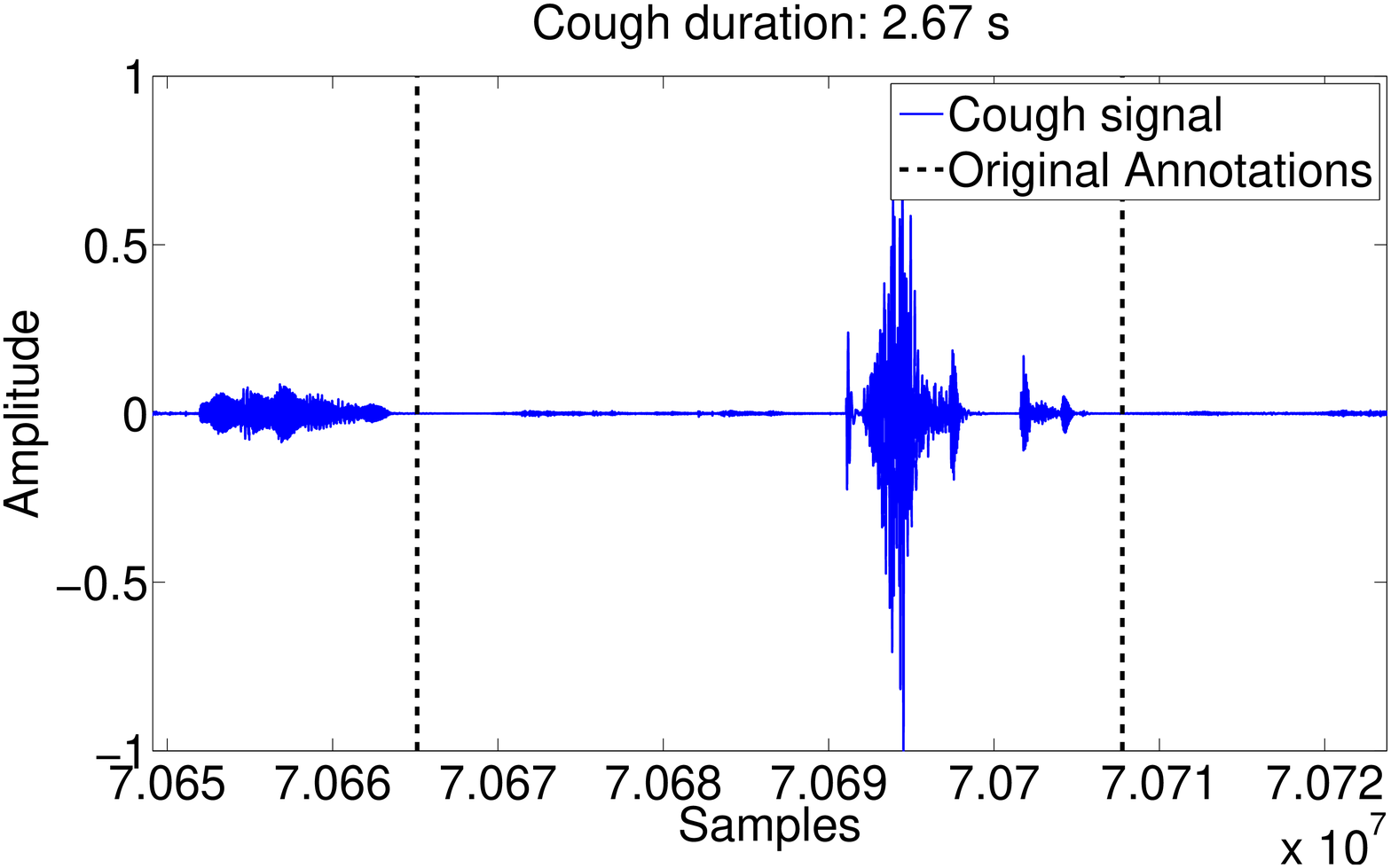}}
    \subfloat{\includegraphics[width=0.5\linewidth]{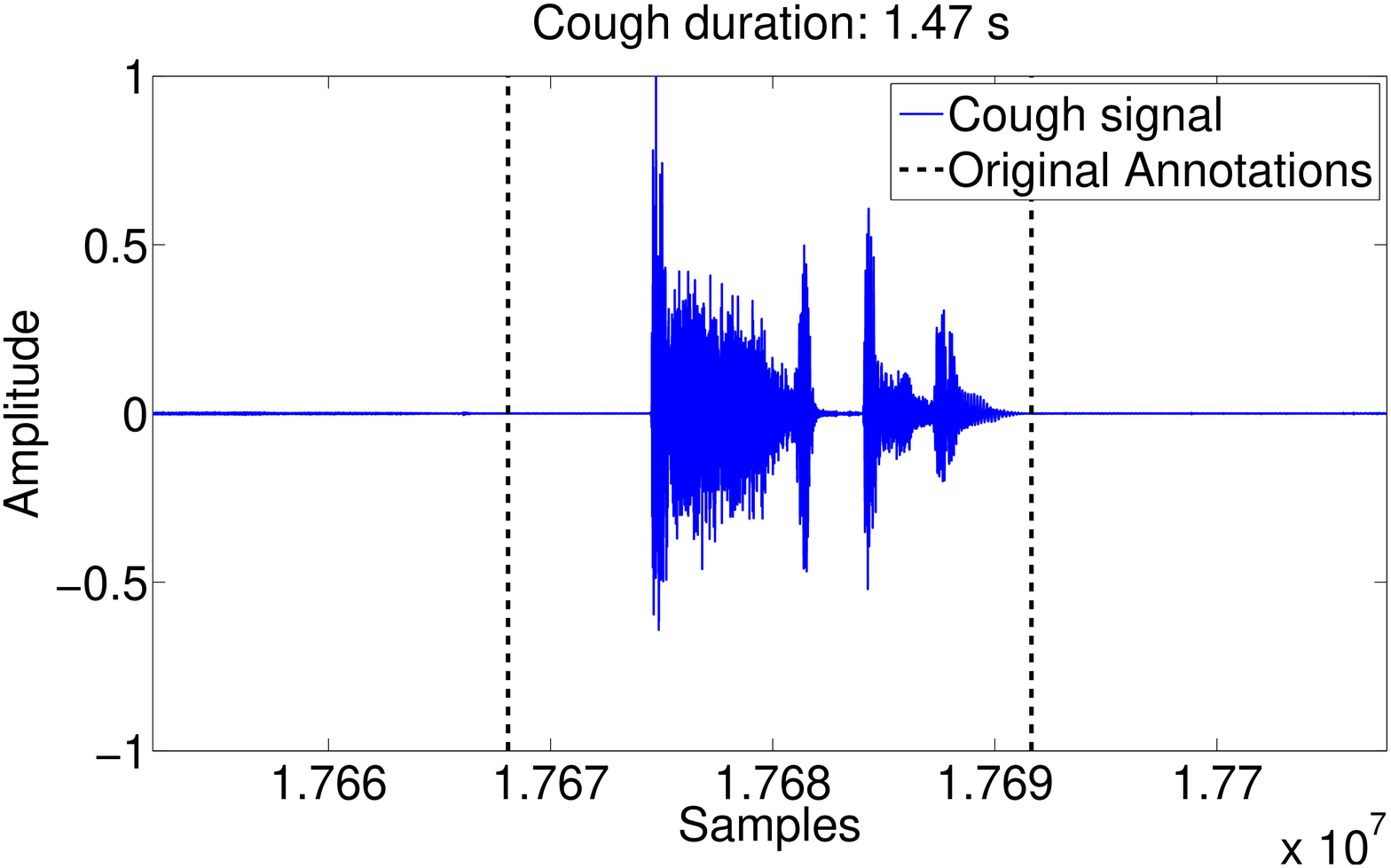}}
    \caption{Examples of inaccurate annotations. Dashed black lines in the figures are the original annotations.}
    \label{fig:Inconsistent} 
\end{figure}

Following an initial analysis of the annotations, it was found that a number of issues existed relating to the original cough event annotations in the AMI corpus including. 
These include incorrect start and end times, and successive coughs annotated as single coughs, examples of which are illustrated in Figure \ref{fig:Inconsistent}.

\begin{figure}[h]
    \centering
    \includegraphics[width=0.7\linewidth]{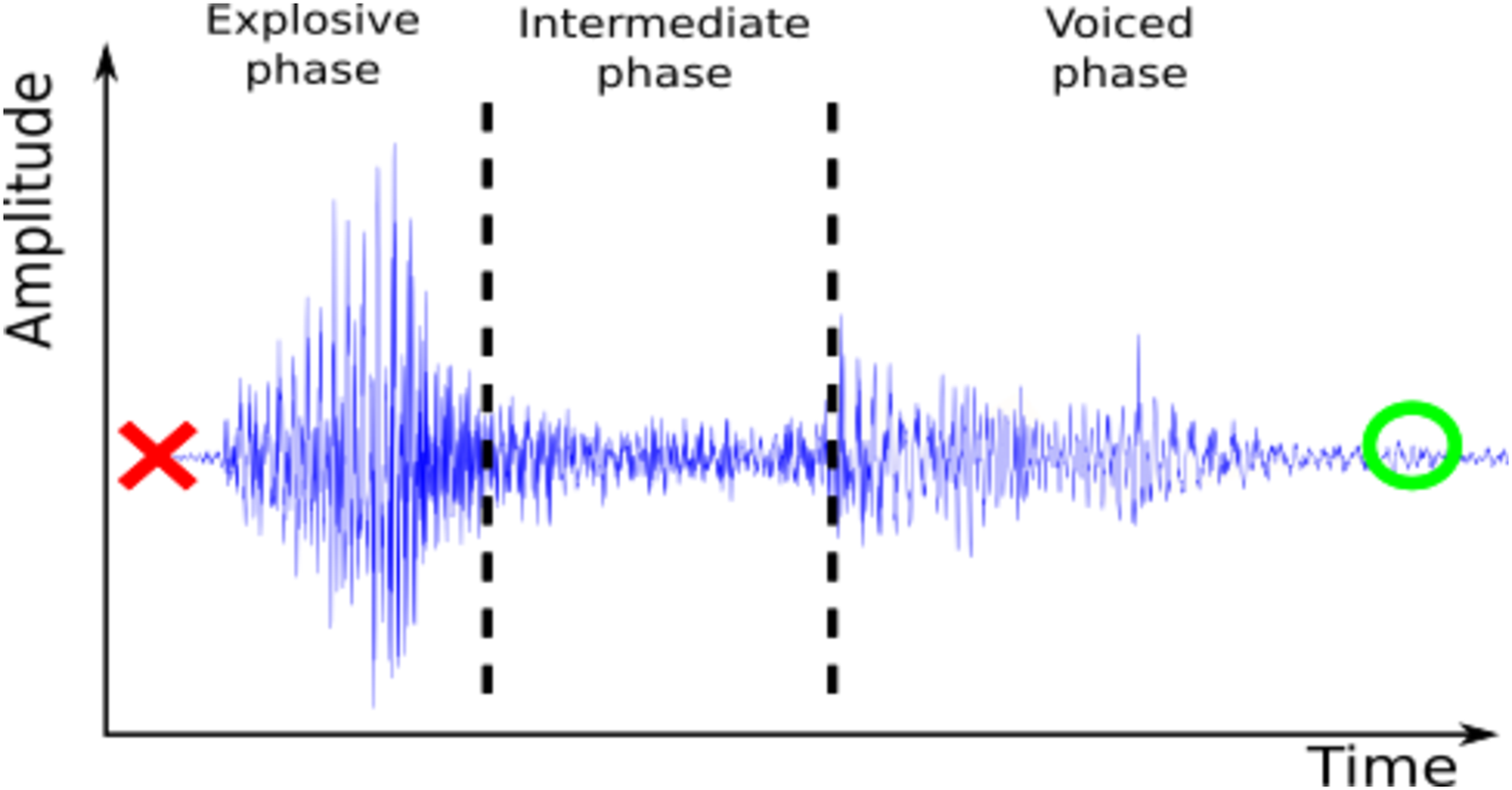}
    \caption{New annotation criteria, red \textbf{``x''} marks the onset/start of the cough, and green \textbf{``o''} marks the end of the cough event.}
    \label{fig:CoughPhasesExample}
\end{figure}

The re-annotation criteria for the cough events is illustrated in Figure \ref{fig:CoughPhasesExample}.

\begin{figure}[b]
    \centering
    \includegraphics[width=0.7\linewidth]{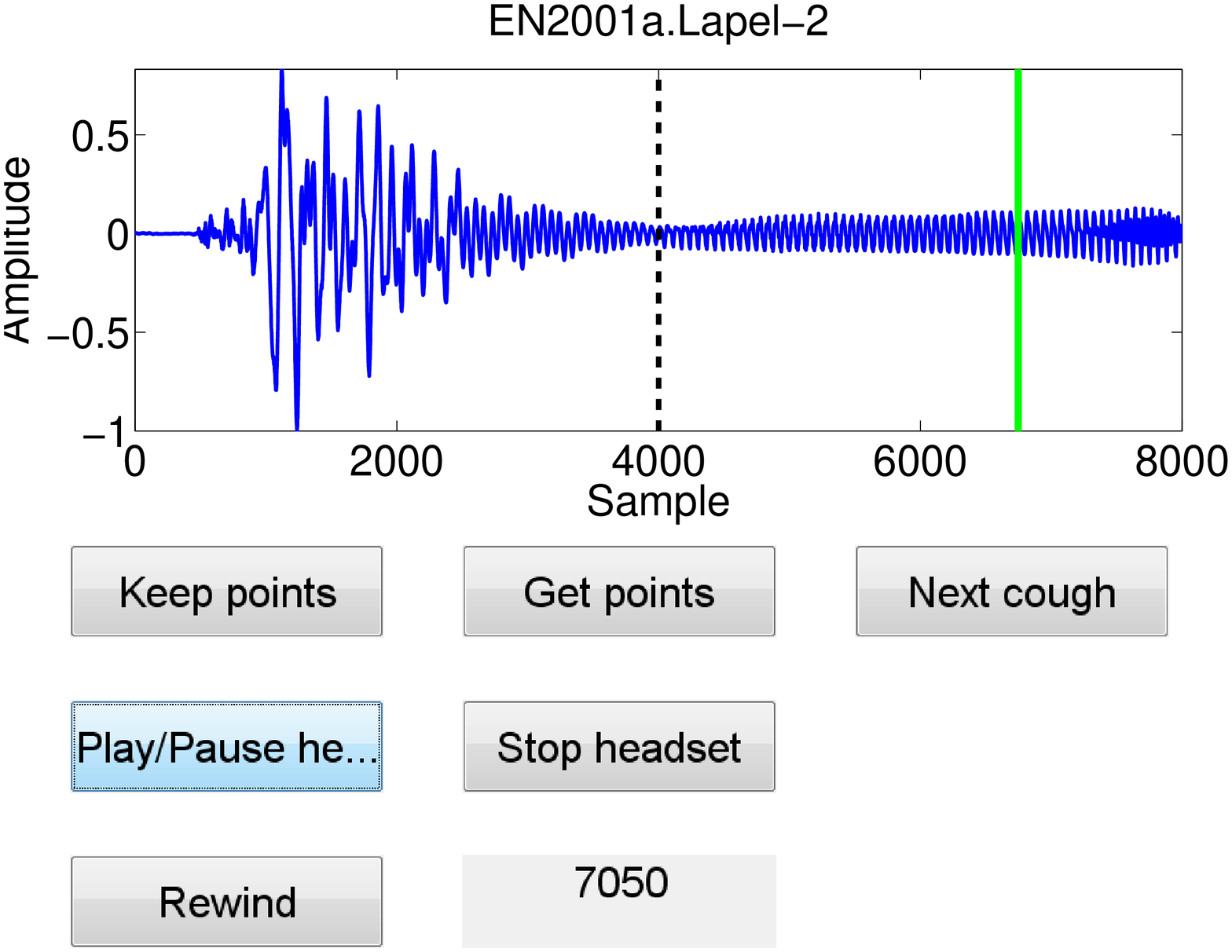}
    \caption{GUI created to carry out the re-annotation procedure of cough events in the AMI corpus.}
    \label{fig:AnnotationTool}
\end{figure}

The task of importing the audio, and the recording of the new time-stamps was carried out using a purpose built GUI, as shown in Figure \ref{fig:AnnotationTool}.

\section{Results}

Following the re-annotation procedure, a new set of time-stamps corresponding to the start and end times of single cough events was produced.
The total number of single cough events before and after the re-annotation procedure is displayed in Table \ref{tab:ReannotatedCoughCount}. 

\begin{table}[h!]
    \caption{Number of cough event annotations}
        \begin{center}
        \begin{tabular}{|c|c|}
        
        \hline
        \multicolumn{2}{|c|}{\textbf{Number of cough annotations}}\\
        \hline
        \textit{Original annotations} & \textit{Re-annotated list} \\
        \hline
        1116 & 1369\\
        \hline
        \end{tabular}
        \label{tab:ReannotatedCoughCount}
    \end{center}
\end{table}

\section{Conclusions}

Justification for the re-annotation of the cough locations in the AMI corpus came following the discovery of a number of issues relating to the original annotations. 
In the area of cough sound analysis and detection the presence of a reliably annotated database will be a useful addition particularly in the area of machine learning. 
This database will be useful for progressing the development and testing of new audio event detection algorithms concerned with cough sounds.
The re-annotated version of the cough location annotations have been made publicly available along with the GUI used in the re-annotation procedure \cite{leamy_2019}. 
A consistent approach to labelling of start and end times of the individual cough events was used, and as a result the potential for the AMI corpus to become a more reliable and reusable source of annotated cough recordings has increased.



\subsection{References}
\bibliographystyle{IEEEtran}
\bibliography{refs}

\end{document}